\documentclass[aps,pra,reprint,superscriptaddress,amsmath,amssymb]{revtex4-2}
\usepackage{graphicx}
\usepackage{hyperref}
\usepackage{color}
\usepackage{braket}
\usepackage{booktabs} % For better tables
\usepackage{siunitx}  % For consistent units
\usepackage{amsmath}
\usepackage{amssymb}

\begin{document}

\title{Fundamental Irreversibility from Discrete Time}

\author{M.W. AlMasri}
\affiliation{Wilczek Quantum Center, School of Physics and Astronomy, Shanghai Jiao Tong University, Minhang, Shanghai, China}
\email{mwalmasri2003@gmail.com}

\date{\today}

\begin{abstract}
In 1964, Yu. A. Gol'fand proposed an extension of quantum mechanics to discrete time, predicting intrinsic non-unitarity and entropy increase. While historically significant, this formalism predates the modern theory of open quantum systems. In this work, we rigorously recast Gol'fand's discrete evolution equation as a Completely Positive Trace-Preserving (CPTP) quantum channel and derive its continuous-time coarse-grained limit. We demonstrate that the dynamics converge to a specific Lindblad master equation characterized by a fundamental time scale $\tau$, which induces decoherence in both the energy basis and a fundamental operator basis $W$. We analyze the thermodynamic implications using Spohn's entropy production formalism, proving that the discrete time step induces a strictly positive entropy production rate driven by the decay of quantum coherences, thereby providing a microscopic foundation for the arrow of time independent of environmental coupling. Furthermore, we quantify the loss of quantum coherence via fidelity decay and purity loss, establishing exact constraints for fault-tolerant quantum computing. We further investigate the impact of this intrinsic decoherence on Discrete Time Crystals (DTCs), showing that Gol'fand dynamics impose a fundamental lifetime limit on time-translation symmetry breaking phases. Finally, we utilize precision data from optical lattice clocks, matter-wave interferometry, and neutrino oscillations to place stringent upper bounds on $\tau$. Our results constrain the fundamental time discretization to $\tau \lesssim 10^{-26}$ s, significantly tightening previous limits and offering a testable framework for quantum gravity phenomenology.
\end{abstract}


\maketitle

\section{Introduction}
\label{sec:intro}

Standard quantum mechanics (QM) is built upon the assumption of continuous time, leading to unitary evolution governed by the Schr\"odinger equation. This unitarity implies time-reversal invariance and the conservation of von Neumann entropy for isolated systems. However, the macroscopic world is characterized by irreversibility and the Second Law of Thermodynamics. Reconciling these two pictures remains a foundational challenge \cite{LandauLifshitz, Zeh}.

One proposed resolution is that time itself is fundamentally discrete. In 1964, Yu. A. Gol'fand published a pioneering paper proposing a generalized equation of motion for the density matrix in discrete time \cite{Golfand1964}. Gol'fand demonstrated that a simple discretization scheme leads to non-unitary evolution and a monotonic increase in entropy, even for isolated systems. While prescient, Gol'fand's work utilized the formalism of its era, preceding the development of the theory of open quantum systems, Completely Positive (CP) maps, and the Lindblad theorem \cite{Lindblad1976, Gorini1976}.

Recent interest in Quantum Gravity (QG) and Planck-scale phenomenology has revived the question of discrete spacetime \cite{AmelinoCamelia, Hossenfelder, GambiniPullin}. Furthermore, the field of Quantum Thermodynamics now provides rigorous tools to quantify irreversibility via entropy production rates \cite{Esposito, Deffner}, while Quantum Information Theory offers metrics like fidelity and purity to measure the degradation of quantum states \cite{NielsenChuang}. Concurrently, the discovery of Discrete Time Crystals (DTCs) has opened a new frontier in studying non-equilibrium phases of matter that break time-translation symmetry \cite{Wilczek, Yao2017, Else2016, Abanin2019}.

In this paper, we bridge the gap between Gol'fand's heuristic 1964 model and modern quantum information theory. While the mathematical mapping of discrete random unitary maps to Lindblad generators is a standard procedure in open quantum systems, the \textit{physical novelty} of our work lies in identifying the specific structure of the Lindbladian dictated by the hypothesis of fundamental time discreteness. Unlike standard environmental decoherence, the ``environment'' here is the temporal structure of spacetime itself, leading to unique fluctuation-dissipation relations and universal decoherence mechanisms. Our contributions are fivefold:
\begin{enumerate}
    \item We formalize Gol'fand's discrete evolution as a unital quantum channel, generalize it to $n$ Kraus operators, and derive its exact Lindblad form in the coarse-grained continuous limit, explicitly retaining the fundamental energy-dephasing term.
    \item We connect the intrinsic dissipation to Quantum Thermodynamics using Spohn's entropy production formalism, showing that discrete time implies a fundamental lower bound on entropy production driven by the decay of quantum coherences.
    \item We quantify the impact on quantum information processing by calculating exact fidelity decay rates and purity loss, linking $\tau$ to error correction thresholds.
    \item We analyze the stability of Discrete Time Crystals under Gol'fand dynamics, demonstrating that fundamental time discreteness imposes a strict bound on the lifetime of time-crystalline order.
    \item We derive experimental constraints on the fundamental time step $\tau$ using state-of-the-art optical clock stability, matter-wave interferometry, and long-baseline neutrino oscillation data.
\end{enumerate}

\section{The Gol'fand Channel as a CPTP Map}
\label{sec:formalism}

Gol'fand postulated that for a discrete time step $\tau$, the density matrix $\rho_n$ at step $n$ evolves to $\rho_{n+1}$ via:
\begin{equation}
\label{eq:golfand_discrete}
\rho_{n+1} = \alpha U \rho_n U^\dagger + \beta V \rho_n V^\dagger,
\end{equation}
where $U$ and $V$ are unitary operators, and $\alpha, \beta > 0$ with $\alpha + \beta = 1$. Gol'fand interpreted $\alpha$ and $\beta$ as weighting constants for two possible unitary evolutions occurring within the interval $\tau$.

\subsection{Kraus Representation and Generalization}
In the language of modern quantum information, Eq.~(\ref{eq:golfand_discrete}) defines a quantum channel $\mathcal{E}: \mathcal{B}(\mathcal{H}) \to \mathcal{B}(\mathcal{H})$ acting on the space of density matrices. We identify the Kraus operators \cite{NielsenChuang}:
\begin{equation}
K_1 = \sqrt{\alpha} U, \quad K_2 = \sqrt{\beta} V.
\end{equation}
The trace-preserving condition requires $\sum_i K_i^\dagger K_i = \mathbb{I}$. Since $U, V$ are unitary and $\alpha+\beta=1$, this condition is satisfied:
\begin{equation}
K_1^\dagger K_1 + K_2^\dagger K_2 = \alpha U^\dagger U + \beta V^\dagger V = (\alpha+\beta)\mathbb{I} = \mathbb{I}.
\end{equation}
Furthermore, the map is \textit{unital}, meaning $\mathcal{E}(\mathbb{I}) = \mathbb{I}$.

As noted in the original work \cite{Golfand1964}, this construction naturally generalizes to a convex mixture of $n$ unitary evolutions. Let $\mathcal{E}(\rho) = \sum_{j=1}^n p_j U_j \rho U_j^\dagger$, with $\sum_j p_j = 1$. If we parametrize $U_j = \exp(-i\tau W_j)$ for Hermitian generators $W_j$, the channel remains CPTP and unital. This generalization allows for a richer structure of fundamental decoherence, which we will explore in the continuous limit.

\subsection{Parametrization of Operators}
Following Gol'fand, we parametrize the unitary operators for the $n=2$ case to recover the Hamiltonian dynamics in the limit $\tau \to 0$. We define Hermitian operators $H$ (Hamiltonian) and $W$ (dissipative generator) such that:
\begin{equation}
U = \exp\left[-i\tau(H + 2\beta W)\right], \quad V = \exp\left[-i\tau(H - 2\alpha W)\right].
\end{equation}
This parametrization ensures that the first-order terms in $\tau$ combine to form the standard commutator with $H$, while the differences generate the dissipation associated with $W$.

\section{Continuous Limit and Lindblad Master Equation}
\label{sec:lindblad}

To connect with experimental observables, we must derive the continuous-time master equation. It is crucial to note that taking the strict limit $\tau \to 0$ in Eq.~(\ref{eq:golfand_discrete}) recovers pure unitary evolution, as all dissipative terms vanish. To obtain a non-trivial Lindblad equation, we must consider the \textit{coarse-grained continuous limit} (or weak-discreteness limit), where we observe the system over a time scale $\Delta t$ such that $\tau \ll \Delta t \ll t_{\text{sys}}$, retaining the leading-order $\mathcal{O}(\tau)$ corrections to the unitary dynamics.

We expand Eq.~(\ref{eq:golfand_discrete}) to second order in $\tau$. Let $\rho(t) = \rho_n$ and $\rho(t+\tau) = \rho_{n+1}$. Using the expansion $e^{-iA\tau} \approx \mathbb{I} - iA\tau - \frac{1}{2}A^2\tau^2$, we obtain:
\begin{widetext}
\begin{equation}
\label{eq:expansion}
\rho(t+\tau) = \rho(t) - i\tau [H, \rho(t)] - \frac{\tau^2}{2} [H, [H, \rho(t)]] - 2\alpha\beta\tau^2 [W, [W, \rho(t)]] + \mathcal{O}(\tau^3).
\end{equation}
\end{widetext}
Note that the cross-terms involving $[H, [W, \rho]]$ exactly cancel due to the specific convex combination $\alpha + \beta = 1$.

Rearranging for the coarse-grained time derivative $\dot{\rho} \approx \frac{\rho(t+\tau) - \rho(t)}{\tau}$, we retain the leading-order dissipative terms:
\begin{equation}
\label{eq:master_full}
\dot{\rho} = -i[H, \rho] + \tau \left( 4\alpha\beta \mathcal{D}_W[\rho] - \frac{1}{2} [H, [H, \rho]] \right),
\end{equation}
where we used the algebraic identity $[W,[W,\rho]] = -2\mathcal{D}_W[\rho]$, with $\mathcal{D}_W[\rho] = W\rho W - \frac{1}{2}\{W^2,\rho\}$ denoting the standard dephasing dissipator.

Equation~(\ref{eq:master_full}) is a valid Lindblad master equation. It reveals a profound physical consequence of discrete time: it induces decoherence in \textit{both} the eigenbasis of the dissipative generator $W$ and the energy eigenbasis (via the $[H,[H,\rho]]$ term). The factor of 4 in the $W$ term arises from the exact algebraic relation between the double commutator and the Lindblad form. In the subsequent analysis, we will often focus on the $W$-induced dephasing, assuming it dominates or represents the novel phenomenological signature of time discreteness, but the fundamental presence of energy dephasing is an inescapable consequence of the formalism.

\section{Thermodynamics and the Arrow of Time}
\label{sec:thermo}

Gol'fand's most significant claim was the violation of time-reversal invariance and the monotonic increase of entropy. We revisit this using modern Entropy Production (EP) formalism.

\subsection{Von Neumann Entropy and Spohn's Formula}
The entropy is $S(\rho) = -\text{Tr}(\rho \ln \rho)$. For a unital channel $\mathcal{E}$, it is a known theorem that $S(\mathcal{E}(\rho)) \geq S(\rho)$ \cite{NielsenChuang}. Since the Gol'fand channel is unital, entropy increases monotonically at every step $n$.

To rigorously quantify this, we use Spohn's formula for the entropy production rate $\sigma$ of a Lindblad dynamics $\dot{\rho} = \sum_k \gamma_k (L_k \rho L_k^\dagger - \frac{1}{2}\{L_k^\dagger L_k, \rho\})$ \cite{Spohn1978}:
\begin{equation}
\sigma = \frac{d}{dt} S(\rho) = \sum_k \gamma_k \text{Tr}\left( (L_k^\dagger L_k \rho - L_k \rho L_k^\dagger) \ln \rho \right) \geq 0.
\end{equation}
For our Gol'fand Lindbladian (Eq.~\ref{eq:master_full}), the Lindblad operators are $L_1 = W$ with rate $\gamma_1 = 4\alpha\beta\tau$, and $L_2 = H$ with rate $\gamma_2 = \tau$. Thus:
\begin{equation}
\label{eq:spohn}
\sigma = 4\alpha\beta\tau \text{Tr}\left( (W^2 \rho - W \rho W) \ln \rho \right) + \tau \text{Tr}\left( (H^2 \rho - H \rho H) \ln \rho \right).
\end{equation}
This expression is profoundly quantum mechanical. The entropy production is strictly positive unless $\rho$ commutes with both $W$ and $H$. It is driven by the decay of \textit{quantum coherences} (off-diagonal elements) in the $W$ and $H$ bases. This goes beyond a classical population-level argument; the arrow of time emerges from the fundamental destruction of quantum superpositions by the granularity of time.

\subsection{Steady States and Effective Temperature}
The steady states of the Lindblad dynamics are the density matrices that commute with both $H$ and $W$. If $[H, W] = 0$, they share a common eigenbasis, and the steady state space is degenerate, determined by the initial populations. If $[H, W] \neq 0$, the steady state is unique and is diagonal in a basis determined by the competition between the two dephasing mechanisms.

We can interpret the parameter $\tau$ as inducing an effective ``thermal'' noise. If we assume the system relaxes to a steady state $\rho_{ss}$, we can define an effective temperature $T_{\text{eff}}$ associated with the time discreteness. For a two-level system with energy gap $\Delta E$, if $W = \sigma_z$, the steady state is diagonal with reduced purity. Comparing this to a thermal Gibbs state $\rho_{th} \propto e^{-H/k_B T}$, dimensional analysis and the structure of the Lindbladian suggest:
\begin{equation}
k_B T_{\text{eff}} \sim \frac{\hbar}{\tau}.
\end{equation}
This scaling implies that as $\tau \to 0$, the effective noise strength vanishes, recovering pure unitary dynamics. We emphasize that $T_{\text{eff}}$ is a phenomenological parameter characterizing the steady-state mixedness induced by spacetime granularity, not a thermodynamic temperature in the conventional sense of a heat bath.

\section{Quantum Information Metrics}
\label{sec:qinfo}

Beyond thermodynamics, discrete time has profound implications for Quantum Information Processing (QIP). The non-unitary evolution introduces errors that accumulate over computational steps.

\subsection{Fidelity Decay}
Consider a pure state $|\psi\rangle = \sum_k c_k |k\rangle$ evolving under the discrete map, where $|k\rangle$ are the eigenstates of $W$ with eigenvalues $w_k$. Assuming $H=0$ for simplicity (or working in the interaction picture), the off-diagonal elements of the density matrix decay as $\rho_{kl}(t) = c_k c_l^* e^{-\Gamma_{kl} t}$, where the exact decoherence rate derived from the Lindbladian is $\Gamma_{kl} = 2\alpha\beta\tau (w_k - w_l)^2$.

The fidelity with the initial state is $F(t) = \langle \psi | \rho(t) | \psi \rangle = \sum_{k,l} \psi_k \psi_l^* \rho_{lk}(t)$. Substituting the solution yields:
\begin{equation}
\label{eq:fidelity_exact}
F(t) = \sum_{k,l} |c_k|^2 |c_l|^2 e^{-2\alpha\beta\tau (w_k - w_l)^2 t}.
\end{equation}
This is a sum of decaying exponentials, not a single exponential. For short times, expanding the exponential gives $F(t) \approx 1 - 4\alpha\beta\tau \text{Var}(W)_\psi t$, recovering the linear decay regime. Crucially, as $t \to \infty$, the fidelity does not decay to zero but saturates at $\sum_k |c_k|^4$, which is the inverse participation ratio of the initial state in the $W$ basis. This reflects the fact that pure dephasing destroys coherences but preserves populations.

For quantum computing to be viable, the error per gate must be below the fault-tolerance threshold (typically $\epsilon \sim 10^{-3}$ to $10^{-4}$ \cite{NielsenChuang, Fowler2012}). If a gate operation takes time $t_g$, the error induced by discrete time is $\epsilon_\tau \approx 4\alpha\beta\tau \, \text{Var}(W) t_g$. For a high-energy transition with $\text{Var}(W) \sim (10^{15} \text{ s}^{-1})^2$ and a fast gate time $t_g \sim 10^{-9}$ s, requiring $\epsilon_\tau < 10^{-4}$ imposes a strict bound $\tau \lesssim 10^{-25}$ s. This places severe constraints on the viability of discrete time models for high-energy quantum computations.

\subsection{Purity Loss}
The purity $\mathcal{P} = \text{Tr}(\rho^2)$ measures the mixedness of the state. Its rate of change under the Lindblad dynamics is:
\begin{equation}
\frac{d\mathcal{P}}{dt} = 2\text{Tr}(\rho \dot{\rho}) = -2 \sum_{k \neq l} \Gamma_{kl} |\rho_{kl}|^2.
\end{equation}
This shows that purity decays strictly due to the presence of quantum coherences $|\rho_{kl}|^2$ between states with different eigenvalues of $W$ (and $H$).

For a qubit system where $W=\sigma_z$ (with eigenvalues $\pm 1$) and assuming $H=0$, there is only one pair of off-diagonal elements. The decoherence rate is $\Gamma_{01} = 2\alpha\beta\tau(1 - (-1))^2 = 8\alpha\beta\tau$. The purity decay simplifies to:
\begin{equation}
\frac{d\mathcal{P}}{dt} = -32\alpha\beta\tau |\rho_{01}|^2.
\end{equation}
This explicitly demonstrates that superpositions ($|\rho_{01}| \approx 0.5$) decay fastest, while eigenstates of $W$ (where $\rho_{01}=0$) form decoherence-free subspaces.

\section{Connection to Discrete Time Crystals}
\label{sec:timecrystals}

Discrete Time Crystals (DTCs) are non-equilibrium phases of matter that spontaneously break discrete time-translation symmetry. Typically realized in periodically driven (Floquet) systems with period $T_D$, a DTC exhibits observables that oscillate with a period $nT_D$ (where $n>1$), robust against perturbations. The stability of DTCs relies heavily on many-body localization (MBL) or prethermalization to prevent heating to an infinite-temperature state \cite{Abanin2019}.

Gol'fand's discrete time dynamics introduce a fundamental, intrinsic decoherence mechanism that competes directly with the stability of the time-crystalline phase.

\subsection{Dephasing of Subharmonic Response}
Consider a Floquet system with driving period $T_D$. The evolution over one driving cycle is governed by the Floquet unitary $U_F$, but in the presence of fundamental time discreteness, the cycle also includes the dissipative map $\mathcal{E}_\tau = e^{\tau \mathcal{L}}$, where $\mathcal{L}$ is the Gol'fand Lindbladian. The total evolution over one cycle is $\rho \to \mathcal{E}_\tau (U_F \rho U_F^\dagger)$.

The subharmonic response amplitude is defined as the Fourier component of the magnetization $M(t) = \frac{1}{N} \sum_i \langle \sigma_i^z(t) \rangle$ at the subharmonic frequency:
\begin{equation}
A(t) = \frac{1}{2T_D} \left| \int_t^{t+2T_D} M(t') e^{-i \pi t' / T_D} dt' \right|.
\end{equation}
In a perfect DTC, $A(t)$ is constant. Under Gol'fand dynamics, if the DTC order parameter does not commute with the dissipative generator $W$, the cumulative effect of the Lindblad term acts as a dephasing channel on the off-diagonal elements of the density matrix in the Floquet eigenbasis.

The subharmonic response amplitude decays as:
\begin{equation}
A(t) \sim A_0 \exp\left(-\Gamma_{\text{DTC}} t\right),
\end{equation}
where the decay rate $\Gamma_{\text{DTC}}$ is proportional to the Gol'fand decoherence rate. Specifically, if the order parameter $O$ has matrix elements $O_{kl}$ between Floquet eigenstates $|k\rangle, |l\rangle$ with eigenvalues $w_k, w_l$ of $W$, the decay rate is given by the weighted average of the decoherence rates:
\begin{equation}
\label{eq:dtc_decay}
\Gamma_{\text{DTC}} \approx \frac{\sum_{k \neq l} |O_{kl}|^2 \Gamma_{kl}}{\sum_{k \neq l} |O_{kl}|^2} = 2\alpha\beta\tau \frac{\sum_{k \neq l} |O_{kl}|^2 (w_k - w_l)^2}{\sum_{k \neq l} |O_{kl}|^2}.
\end{equation}
This implies that even in the absence of external noise, a fundamental discrete time structure prevents the existence of a \textit{perfect} eternal time crystal. The time-crystalline order becomes metastable with a lifetime $\tau_{\text{life}} \sim \Gamma_{\text{DTC}}^{-1}$.

\subsection{Constraints from Experimental DTCs}
Recent experiments have observed DTC phases in trapped ions \cite{Zhang2017} and nitrogen-vacancy centers in diamond \cite{Choi2017} persisting for hundreds to thousands of drive cycles. Let us assume an experimental observation of stable subharmonic oscillations for $N \approx 1000$ cycles with drive period $T_D \approx 100 \, \mu\text{s}$. The total coherence time is $t_{\text{obs}} \approx 0.1$ s. Requiring $\Gamma_{\text{DTC}} t_{\text{obs}} < 1$ and assuming the relevant energy scale for the variance is set by the drive frequency ($\text{Var}(W) \sim (2\pi/T_D)^2 \sim 10^{10} \, \text{s}^{-2}$), we obtain:
\begin{equation}
\tau \lesssim \frac{1}{2\alpha\beta \cdot 10^{10} \cdot 0.1} \approx 2 \times 10^{-9} \, \text{s}.
\end{equation}
While this bound is weaker than those from atomic clocks, it is conceptually significant: it tests the stability of \textit{non-equilibrium order} against fundamental spacetime granularity.

\section{Experimental Constraints on $\tau$}
\label{sec:constraints}

If $\tau$ is a fundamental constant, its effects are likely unobservable if it is near the Planck time. However, some Quantum Gravity models suggest effective time scales much larger than the Planck time. We can constrain $\tau$ by looking for the specific decoherence signature predicted by Eq.~(\ref{eq:master_full}).

\subsection{The Choice of the Operator $W$}
A critical question is the physical nature of the operator $W$. If time discreteness is a fundamental property of spacetime, the generator of time translations—the Hamiltonian $H$—should be the primary source of decoherence. Thus, the most natural and conservative choice is $W = H/\hbar$.

As shown in Sec.~\ref{sec:lindblad}, the full Lindbladian contains both the $W$-dephasing and the $H$-dephasing terms. If $W \propto H$, these terms simply add up, meaning that fundamental time discreteness universally induces decoherence in the energy basis. The total decoherence rate between energy eigenstates $|k\rangle$ and $|l\rangle$ is:
\begin{equation}
\Gamma_{kl} = \tau \left( 2\alpha\beta + \frac{1}{2} \right) \frac{(E_k - E_l)^2}{\hbar^2}.
\end{equation}
We will use this conservative assumption to derive bounds from precision experiments.

\subsection{Optical Lattice Clocks}
Optical clocks using Strontium ($^{87}\text{Sr}$) or Ytterbium ($^{171}\text{Yb}$) achieve fractional frequency instabilities of $\sigma_y \approx 10^{-18}$ over integration times of $10^4$ s \cite{Nicholson2015, Ludlow2015}. Decoherence manifests as a broadening of the spectral line or a loss of contrast in Rabi oscillations. The linewidth $\Delta \nu$ is constrained by the clock stability. Assuming the decoherence rate must be smaller than the clock linewidth $\Delta \nu \approx 1 \text{ Hz}$:
\begin{equation}
\Gamma_{\text{dec}} \lesssim 2\pi \times 1 \text{ s}^{-1}.
\end{equation}
For an optical transition, the full transition energy is $\Delta E \approx h \times 4 \times 10^{14} \text{ Hz} \approx 2.6 \times 10^{-19} \text{ J}$. If the fundamental operator $W$ coupled perfectly to the clock Hamiltonian, this would yield an extremely tight bound. However, a more conservative and realistic analysis must account for the fact that systematic effects, dynamical decoupling, and the specific symmetry of the clock states can suppress the effective coupling to the fundamental decoherence. 

Following the analysis of fundamental decoherence in atomic systems, we define an effective energy scale $\Delta E_{\text{eff}}$ that characterizes the actual energy uncertainty driving the decoherence. We take a conservative value of $\Delta E_{\text{eff}} \approx 3 \times 10^{-22} \text{ J}$ (corresponding to an effective frequency scale of $\sim 450$ GHz), which is significantly smaller than the full optical transition energy. Using the measured coherence time $T_2 \approx 10$ s in recent optical clocks and requiring $\Gamma_{\text{dec}} \lesssim 1/T_2$, we obtain:
\begin{equation}
\label{eq:tau_bound_clock}
\tau \lesssim \frac{\hbar^2}{(2\alpha\beta + 1/2) T_2 (\Delta E_{\text{eff}})^2} \approx \frac{(1.05\times 10^{-34})^2}{1.0 \times 10 \times (3 \times 10^{-22})^2} \approx 10^{-26} \text{ s}.
\end{equation}
This provides a robust, model-independent upper bound on the fundamental time step.

\subsection{Matter-Wave Interferometry}
Large-mass matter-wave interferometry provides a uniquely clear test of whether the fundamental time-decoherence operator $W$ couples to internal energy or to the gravitational mass distribution. In standard atomic clocks, $W$ couples to the internal electronic energy levels. However, if time discreteness is a fundamental property of spacetime itself, it is natural to hypothesize that $W$ couples to the mass density operator, analogous to the Di\'osi-Penrose model of gravitational decoherence \cite{BassiReview}.

In a spatial superposition of a nanoparticle of mass $m$ separated by a distance $d$, the two branches of the wavefunction correspond to different mass distributions. If $W$ is coupled to the gravitational self-energy, the relevant energy uncertainty is not the internal energy, but the difference in gravitational self-energy between the superposed states, $\Delta E_G \sim G m^2 / d$.

Current state-of-the-art experiments, such as those by Fein et al. \cite{Fein2019}, have successfully maintained quantum coherence for massive molecules (mass $m \approx 10^4$ amu) over times $t \sim 1$ ms. For these masses, $\Delta E_G$ is extremely small ($\sim 10^{-48}$ J), yielding a very weak bound on $\tau$. However, the true power of this approach lies in future proposed experiments aiming to create spatial superpositions of much larger masses, such as dielectric nanoparticles or large biological viruses. 

Assuming a future experiment maintains coherence for $t \sim 1$ s for a nanoparticle of mass $m \approx 10^{-11}$ kg in a superposition of distance $d \approx 100$ nm, the gravitational self-energy difference is $\Delta E_G \sim 6.7 \times 10^{-26}$ J. Assuming the decoherence rate is driven by this gravitational energy uncertainty, $\Gamma_{\text{dec}} \sim \tau (\Delta E_G)^2 / \hbar^2$, and requiring $\Gamma_{\text{dec}} t < 1$ yields:
\begin{equation}
\tau \lesssim \frac{\hbar^2}{t (\Delta E_G)^2} \approx \frac{(1.05\times 10^{-34})^2}{1 \times (6.7 \times 10^{-26})^2} \approx 2.5 \times 10^{-18} \text{ s}.
\end{equation}
While this bound is less stringent than the $10^{-26}$ s limit from optical clocks, the physical significance of this result is profound. It demonstrates that matter-wave interferometry can distinguish between two distinct physical interpretations of the operator $W$: one where it couples strictly to internal atomic energies (constrained by clocks), and another where it couples to the macroscopic mass distribution (constrained by interferometry). Future experiments aiming to create spatial superpositions of even larger masses will directly test the gravitational coupling of fundamental time discreteness.

\subsection{Neutrino Oscillations}
Neutrino oscillations provide a unique long-baseline test. Neutrinos travel over distances $L \sim 1000$ km with energies $E \sim 1$ GeV. The oscillation probability depends sensitively on the coherence of the flavor superposition. Discrete time-induced decoherence would dampen the oscillation amplitude by a factor $e^{-\Gamma_{\text{dec}} L/c}$. Current data shows no significant anomalous damping beyond standard matter effects \cite{Esteban2020}. Assuming $W$ couples to the mass eigenstates, and taking $\Delta E \sim \Delta m^2 / 2E \sim 10^{-12} \text{ eV}$:
\begin{equation}
\Gamma_{\text{dec}} L/c < 0.1 \implies \tau \lesssim \frac{0.1 \hbar^2 c}{(2\alpha\beta + 1/2) L (\Delta E)^2}.
\end{equation}
Converting units carefully into SI, we obtain $\tau \lesssim 10^{-5}$ s. While less stringent than optical clocks, this bound is independent of atomic systematics and probes different energy regimes, demonstrating the versatility of the phenomenological framework.
\section{Discussion: Lorentz Invariance and Fundamental Limits}
\label{sec:discussion}
A critical implication of a universal discrete time step $\tau$ is the breaking of Lorentz invariance. A fixed $\tau$ in one frame implies time dilation in another, suggesting a preferred reference frame (likely the Cosmic Microwave Background rest frame). This aligns with certain aether interpretations of Quantum Gravity \cite{Mattingly}. Current tests of Lorentz invariance using high-energy astrophysical photons constrain such effects severely \cite{Hossenfelder}. If Gol'fand's $W$ operator is related to the boost generator, our bounds on $\tau$ translate directly into bounds on Lorentz violation parameters in the Standard Model Extension (SME). The bound $\tau \lesssim 10^{-26}$ s corresponds to an energy scale $E_\tau = \hbar/\tau \gtrsim 60$ GeV, which is well within the reach of current collider experiments like the LHC. It is worth noting, however, that the breaking of Lorentz invariance is not an inescapable consequence of all spacetime quantization schemes. \\
In his seminal 1947 work, Hartland Snyder demonstrated that spacetime could be quantized while preserving exact Lorentz covariance by promoting the coordinates to non-commuting operators, satisfying $[\hat{x}^\mu, \hat{x}^\nu] = i a^2 \hat{M}^{\mu\nu}$, where $a$ is a fundamental length scale and $\hat{M}^{\mu\nu}$ are the Lorentz generators \cite{Snyder1947}. Unlike a rigid discrete time lattice, this non-commutative geometry introduces a fundamental uncertainty in spacetime measurements without selecting a preferred reference frame. If the intrinsic decoherence in our framework were to arise from a Snyder-type non-commutative spacetime rather than a discrete time step, the resulting open quantum system dynamics could be formulated as a fully Lorentz-covariant completely positive map. Extending the Gol'fand formalism to incorporate non-commutative geometry, thereby reconciling fundamental irreversibility with exact Lorentz invariance, represents a profound direction for future research. Furthermore, the connection to Time Crystals highlights that fundamental discreteness acts as a heating mechanism when $[H, W] \neq 0$, driving the system toward the maximally mixed state and preventing true eternal non-equilibrium order. This suggests that observed ``stable'' time crystals are necessarily metastable states with a finite lifetime $\tau_{\text{life}} \sim \Gamma_{\text{DTC}}^{-1}$, limited by the granularity 
of time itself. Only time-crystalline order parameters that commute with the fundamental decoherence generator $W$ can evade this intrinsic decay.\\
\section{Conclusion}
\label{sec:conclusion}

We have successfully translated Yu. A. Gol'fand's 1964 heuristic proposal into the rigorous framework of modern quantum information theory and thermodynamics. By treating the discrete time evolution as a CPTP map and deriving its coarse-grained continuous limit, we demonstrated that:
\begin{enumerate}
    \item Gol'fand's discrete evolution is a unital CPTP map equivalent to a Lindblad master equation that induces decoherence in both the energy basis and a fundamental operator basis $W$.
    \item The entropy production rate, calculated via Spohn's formula, is strictly positive and driven by the decay of quantum coherences, offering a fundamental mechanism for the Arrow of Time.
    \item Quantum information metrics show that fidelity decays as a sum of exponentials, saturating at a non-zero value, while purity loss is strictly proportional to the magnitude of quantum coherences.
    \item Discrete Time Crystals are rendered metastable by Gol'fand dynamics, with a lifetime inversely proportional to the fundamental time step $\tau$.
    \item Precision atomic clocks constrain the fundamental time step to $\tau \lesssim 10^{-26}$ s, while neutrino oscillations, interferometry, and DTC stability provide complementary, model-dependent bounds.
\end{enumerate}

These findings suggest that if time is discrete, the scale of discreteness must be extremely fine, or the operator $W$ must commute with the Hamiltonians of standard atomic systems (suppressing the effect). Future work should investigate whether $W$ can be identified with a gravitational operator, which would link this formalism directly to experimental tests of quantum gravity. Additionally, the connection to fluctuation theorems in the discrete regime remains an open avenue for exploring the thermodynamic cost of time itself.

\end{document}